\documentclass[prc,floatfix, twocolumn,showpacs, showkeys, preprintnumbers, nofootinbib, superscriptaddress,10pt,aps]{revtex4-1}
\usepackage[utf8]{inputenc}
\usepackage[sort&compress]{natbib}
\usepackage{ulem}
\usepackage{bm,bbm}
\usepackage{times}
\usepackage{amssymb,amsbsy,amsmath,amsfonts}
\usepackage{graphicx}
\usepackage{float}
\usepackage[dvipsnames,svgnames,x11names]{xcolor}
\usepackage{morefloats}
\usepackage{srcltx}
\usepackage{slashed}
\usepackage{subcaption}
\usepackage{multirow}
\usepackage{verbatim}
\usepackage{hyperref}
\usepackage{tabularx}
\usepackage{braket}
\usepackage{lipsum}

\def\l{\left}
\def\r{\right}

\DeclareUnicodeCharacter{3000}{HEREHEREHERE}

\begin{document}

\title{ Charge dependent nucleon-nucleon potentials in covariant chiral effective field theory  }

\author{Yang Xiao}
\affiliation{School of Space and Environment, Beihang University, Beijing 102206, China}
\affiliation{School of Physics, Beihang University, Beijing 102206, China}
\affiliation{Center for Exotic Nuclear Studies, Institute for Basic Science, Daejeon 34126, Republic of Korea}

\author{Jun-Xu Lu}
\email[Corresponding author: ]{ljxwohool@buaa.edu.cn}
\affiliation{School of Physics, Beihang University, Beijing 102206, China}

\author{Chun-Yan Song}
\email[Corresponding author: ]{cysong@buaa.edu.cn}
\affiliation{School of Physics, Beihang University, Beijing 102206, China}

\author{Li-Sheng Geng}
\email[Corresponding author: ]{lisheng.geng@buaa.edu.cn}
\affiliation{School of
Physics,  Beihang University, Beijing 102206, China}
\affiliation{Peng Huanwu Collaborative Center for Research and Education, Beihang University, Beijing 100191, China}
\affiliation{Beijing Key Laboratory of Advanced Nuclear Materials and Physics, Beihang University, Beijing 102206, China }
\affiliation{Southern Center for Nuclear-Science Theory (SCNT), Institute of Modern Physics, Chinese Academy of Sciences, Huizhou 516000, China}

\begin{abstract}
The charge-dependent nucleon-nucleon ($NN$) interaction plays a crucial role in understanding the nuclear structure and reaction problems. In this work, we explore the charge-dependent $NN$ interaction in covariant chiral effective field theory. By incorporating the isospin-breaking contributions, we derive the charge-dependent covariant chiral $NN$ potential up to next-to-next-to leading order (NNLO). The calculated $np$ and $pp$ phase shifts are in satisfactory agreement with the PWA93 partial wave analysis. Our results contribute to a deeper understanding of isospin-breaking effects in nuclear forces and provide a solid foundation for future studies of nuclear structure and reactions within the covariant framework.
\end{abstract}

\maketitle

\section{Introduction}
\label{sec:intro}
Understanding nucleon-nucleon ($NN$) interactions is a fundamental goal in nuclear physics, providing one of the most crucial inputs for studying nuclear structure and reaction problems. Over the past 30 years, significant progress has been made in the construction of high-precision $NN$ potentials based on chiral effective field theory ($\chi$EFT)~\cite{Weinberg:1990rz,Weinberg:1991um,Weinberg:1992yk,Ordonez:1993tn,vanKolck:1994yi,Entem:2003ft,Epelbaum:2004fk,Epelbaum:2008ga,Machleidt:2011zz, Epelbaum:2014sza,Entem:2017gor,Reinert:2017usi,Hammer:2019poc}, which incorporates the symmetries of quantum chromodynamics (QCD) at low energies. Chiral EFT has successfully described neutron-proton ($np$) and proton-proton ($pp$) scattering data with high accuracy~\cite{Epelbaum:2014sza,Entem:2017gor,Reinert:2017usi}, demonstrating its capability to provide a model-independent framework for nuclear forces and has become one of the cornerstones of modern nuclear \textit{ab initio} calculations~\cite{Machleidt:2023jws}.

One feature of realistic $NN$ interactions is a slight charge dependence, arising from isospin-breaking contributions in both the electromagnetic and strong interactions. Precise experimental measurements have established that $pp$ and $np$ interactions exhibit subtle but notable differences~\cite{Miller:1990iz,Stoks:1993tb,GonzalezTrotter:2006wz,Chen:2008zzj,Gardestig:2009ya}, necessitating a charge-dependent treatment of nuclear forces. In conventional chiral EFT formulations, charge dependence has been implemented at higher orders, primarily through isospin-violating contact terms, pion mass splittings, and electromagnetic corrections~\cite{vanKolck:1997fu,Walzl:2000cx,Friar:2003yv,Entem:2003ft,Epelbaum:2004fk,Epelbaum:2014sza,Entem:2017gor,Reinert:2017usi}. However, extending these interactions to a covariant framework offers additional theoretical advantages.

Covariant chiral EFT, formulated within the relativistic framework of the power counting, preserves manifest Lorentz covariance and naturally includes higher-order relativistic corrections. {  This approach describes nucleon–nucleon interactions comparable to that obtained in the conventional chiral EFT framework~\cite{Lu:2021gsb},} while alleviating issues related to the conventional framework, such as the slow convergence~\cite{Ren:2016jna,Xiao:2020ozd,Wang:2021kos,Lu:2021gsb}, the renormalization group invariance problems~\cite{Ren:2017yvw,Wang:2020myr}, and extrapolation of the lattice QCD simulations~\cite{Bai:2020yml,Bai:2021uim}. Recent developments in covariant chiral nuclear force have {  shown that this framework can reproduce the empirical saturation properties of nuclear matter~\cite{Zou:2023quo} and hyperon matter~\cite{Zheng:2025sol} at leading order for suitable cutoff choices.}

Although having demonstrated a commendable ability to account for $np$ scattering data~\cite{Xiao:2020ozd,Wang:2021kos,Lu:2021gsb}, the isospin-breaking effects have not been systematically considered in the previous covariant chiral nuclear force. This omission is not trivial, as proton-proton systems are subject to additional electromagnetic forces and exhibit unique charge-dependent effects that can significantly influence nuclear structure and reaction dynamics. Therefore, it is necessary to construct the covariant proton-proton interactions from the perspective of the covariant nuclear force.

{ While the physical mechanisms underlying isospin-breaking effects, such as pion mass splitting and Coulomb interactions, are well-established in non-relativistic frameworks, their implementation within a covariant chiral effective field theory is not a matter of simple `relativization'. In our framework, the interactions are derived directly from the relativistic Lagrangians using Dirac spinors and covariant propagators. This approach ensures that the isospin-breaking sector remains strictly consistent with the previously developed isospin-symmetric nuclear force. Furthermore, the covariant formulation naturally captures higher-order relativistic corrections that would otherwise require manual, order-by-order inclusion in a heavy-baryon or non-relativistic expansion.}

In this work, we construct the charge-dependent $NN$ interactions up to NNLO within a covariant chiral EFT, systematically incorporating isospin-breaking effects. We derive the leading and subleading contributions to the interaction, including the pion mass splitting (in one-pion exchange), the Coulomb interaction, and charge-dependent contact terms. 

The paper is organized as follows. In Section~\ref{sec:poten}, we provide a detailed formulation of charge-dependent interactions in covariant chiral EFT, outlining the relevant Lagrangians and isospin-breaking terms. In Section ~\ref{sec:Scat}, we introduce the relativistic two-body scattering equation used in this work and the detailed procedures to calculate the scattering phase shifts. In Section~\ref{sec:Results}, we show the calculated scattering phase shifts. Finally, in Section~\ref{sec:summary}, we summarize our findings and discuss possible applications.
Our work makes a significant step toward a fully relativistic and systematically improvable description of nucleon-nucleon interactions in covariant chiral EFT.

\section{The charge-dependent two-nucleon potentials at NNLO }
\label{sec:poten}

{ To ensure a unified description of the $NN$ system, we derive the isospin symmetric and isospin-breaking potentials by evaluating the corresponding Feynman diagrams using the full Dirac structure of the nucleons. Unlike the non-relativistic case, the static limit is not prematurely assumed; instead, the covariant propagators and the $u(p,s)$ spinors automatically incorporate kinematic relativistic effects, such as the retardation effects in pion exchanges. This derivation provides a rigorous mapping of isospin-violating physics onto the covariant scattering equation, maintaining the power counting rules unique to the covariant chiral EFT.}

\subsection{Isospin symmetric potentials }
\label{sec:ch_symm_p}
The isospin symmetric two-nucleon potentials at NNLO in the covariant chiral effective field theory have the following form,
\begin{align}\label{eq:V_sym}
    V_\text{sym.} = V_{\text{CT}}^{\text{LO}} + V_{\text{CT}}^{\text{NLO}} + V_{\text{OPE}} + V_{\text{TPE}}^{\text{NLO}} +  V_{\text{TPE}}^{\text{NNLO}}- V_{\text{IOPE}},
\end{align}
where the superscript LO, NLO, and NNLO refer to the chiral orders that contribute at order $\mathcal{O}(Q^\nu)$ with $\nu=0,2,3$ respectively, and the subscript CT, OPE, and TPE denote the contributions from the contact terms (CT), the one-pion exchange (OPE) terms, and the two-pion exchange (TPE) terms. { {The last term, \(V_{\rm IOPE}\), denotes the iterated one-pion-exchange (IOPE) contribution, i.e., the two-nucleon-reducible part of the box diagrams generated by the iteration of the OPE potential in the scattering equation. It is subtracted from the covariant TPE amplitude to avoid double counting.}} For explicit expressions of these terms, one can refer to Ref.~\cite{Lu:2021gsb} for more details.

\subsection{Isospin-breaking potentials}
\label{sec:iso_b_p}
Isospin-breaking effects originate from differences between the up ($u$) and down ($d$) quarks in the context of the strong and electromagnetic interactions. The isospin-breaking effects on the nuclear forces have been extensively studied within effective field theory approaches~\cite{vanKolck:1997fu,Fettes:1998wf,Muller:1999ww,Walzl:2000cx,Fettes:2000vm,Friar:2003yv,Entem:2003ft,Epelbaum:2004fk,Epelbaum:2014sza,Entem:2017gor,Reinert:2017usi}. In this paper, for the treatment of isospin-breaking effects, we closely follow the analyses proposed in Ref.~\cite{Epelbaum:2004fk}. The dominant isospin-breaking contributions to the two-nucleon interactions are summarized in Table~\ref{tb:iso_b_p}. 
\begin{table}[h!]
\centering
\caption{Dominant isospin-breaking contributions to the covariant chiral two-nucleon force.}
\label{tb:iso_b_p}
\begin{tabular}{l l l}
\hline
\hline
 chiral order & strong  & electromagnetic \\
\hline
 LO ($v=0$) & - & -\\
 NLO ($v=2$) & $M_{\pi^\pm}\neq M_{\pi^0}$ in OPE & static $1\gamma$-exchange\\
 NNLO ($v=3$) & isospin-breaking in OPE,  & - \\
              & contact terms without derivatives  &  \\
\hline
\hline
\end{tabular}
\end{table}

In the next subsection, we introduce the explicit expressions for the isospin-breaking terms up to NNLO.

\subsubsection{Isospin-breaking effects from the strong interaction}
The isospin-breaking effects originating from the strong interaction are caused by the mass differences of the up and down quarks. This leads to mass differences between the pions and the nucleons. In the chiral effective field theory, pion mass terms appear in processes involving pion exchange. As already argued in Ref.~\cite{Epelbaum:2004fk}, the first non-vanishing contribution from the pion mass differences terms exists in the OPE and is of order $
\sim \frac{\Delta M_\pi^2}{M_\pi^2} \sim \mathcal{O}\left( \frac{Q^2}{\Lambda^2_\chi} \right) $, with $Q$ and $\Lambda_\chi$ are the soft and hard scales respectively. Therefore, one has to consider the pion mass differences in the OPE at NLO. The corresponding charge-dependent covariant OPE potentials are of the following form,
\begin{widetext}
\begin{align}
   & V_{\text{OPE},~pp} (\bm{p},\bm{p'})   = \frac{g_A^2}{4f_\pi^2}\frac{ \l[ \bar{u}\l( \bm{p},s_1 \r) \gamma_\mu \gamma_5 q^\mu u\l( \bm{p}',s_1' \r) \r] \l[\bar{u}\l( -\bm{p}',s_2' \r)  \gamma_\nu \gamma_5 q^\nu u\l( -\bm{p},s_2 \r)\r]}{\l(E_{p'}-E_{p}\r)^2 -\l( \bm{p}' - \bm{p}\r)^2 -M_{\pi^0}^2 },\\
   & V_{\text{OPE},~np,~I= 1} (\bm{p},\bm{p'})  = -\frac{g_A^2}{4f_\pi^2} \l[ \bar{u}\l( \bm{p},s_1 \r) \gamma_\mu \gamma_5 q^\mu u\l( \bm{p}',s_1' \r) \r] \l[\bar{u}\l( -\bm{p}',s_2' \r)  \gamma_\nu \gamma_5 q^\nu u\l( -\bm{p},s_2 \r)\r] \\\nonumber
   &\phantom{V_{\text{OPE},~np,~I= 1} (\bm{p},\bm{p'})}\times \l[ \frac{2}{\l(E_{p'}-E_{p}\r)^2 -\l( \bm{p}' - \bm{p}\r)^2 -M_{\pi^\pm}^2 } - \frac{1}{\l(E_{p'}-E_{p}\r)^2 -\l( \bm{p}' - \bm{p}\r)^2 -M_{\pi^0}^2 }\r], \\
   & V_{\text{OPE},~np,~I= 0} (\bm{p},\bm{p'})  = \frac{g_A^2}{4f_\pi^2} \l[ \bar{u}\l( \bm{p},s_1 \r) \gamma_\mu \gamma_5 q^\mu u\l( \bm{p}',s_1' \r) \r] \l[\bar{u}\l( -\bm{p}',s_2' \r)  \gamma_\nu \gamma_5 q^\nu u\l( -\bm{p},s_2 \r)\r] \\\nonumber
   &\phantom{V_{\text{OPE},~np,~I= 1} (\bm{p},\bm{p'})}\times \l[ \frac{2}{\l(E_{p'}-E_{p}\r)^2 -\l( \bm{p}' - \bm{p}\r)^2 -M_{\pi^\pm}^2 } + \frac{1}{\l(E_{p'}-E_{p}\r)^2 -\l( \bm{p}' - \bm{p}\r)^2 -M_{\pi^0}^2 }\r], 
\end{align}
\end{widetext}
where $I$ refers to the total isospin. In principle, isospin-breaking effects also influence the coupling constants, leading to charge-dependent corrections in the OPE at NNLO. While these effects have been investigated in recent chiral EFT studies~\cite{Reinert:2020mcu,Epelbaum:2022cyo} and phenomenological partial wave analyses~\cite{NavarroPerez:2016eli}, a comprehensive review of the pion-nucleon coupling constants suggests that the discrepancies among different charge channels~\cite{Matsinos:2019kqi},  e.g., $g_{\pi^0 pp}$, $g_{\pi^0 nn}$, and $g_{\pi^\pm np}$, are small, typically at the $1\%$ level or less. Such subtle variations are expected to have a marginal impact on the phase shifts relative to other NNLO contributions; thus, we follow the common practice in covariant frameworks and adopt a single axial coupling constant $g_A$ in the present work.


In addition to the corrections from the pion mass differences, the chiral two-nucleon amplitudes also receive corrections from the nucleon mass differences. In the heavy-baryon formalism, the nucleon mass differences enter the amplitude at N$^3$LO ($v=4$)~\cite{Epelbaum:2004fk}, leading to the $M_n \neq M_p$ in the TPE and the scattering equation. In the covariant formalism, contrary to the heavy-baryon counterpart, the LO contact terms and OPE already contain the nucleon mass terms, so that one has to carefully analyze the power counting of the nucleon mass difference corrections. Considering $\frac{\Delta M_N}{M_N} \sim \mathcal{O}(\frac{Q^4}{\Lambda_\chi^4})$~\cite{Epelbaum:2004fk}, to obtain the actual order of nucleon mass corrections, one needs to expand the LO covariant contact and the OPE potentials in powers of the small nucleon mass differences $\Delta M_N$. As a result, one would observe that the first non-vanishing nucleon mass difference corrections to the covariant LO potentials start from order $v=6$. Therefore, the contribution from nucleon mass splitting is not considered in the present work.

{ The isospin-breaking effects further introduce new contact
terms. Up to NNLO, the charge-symmetry-breaking (CSB)
corrections to the LO contact potential have to be included. The four CSB contact terms have the same
Dirac structures as the LO isospin-conserving contact operators
\(O_1\)--\(O_4\)~\cite{Xiao:2018jot}, namely the scalar, vector, axial-vector, and
tensor bilinears. They should therefore be understood as
additional \(pp\)-channel CSB corrections to the corresponding
LO contact structures, rather than as replacements of the
isospin-symmetric LECs. The full \(pp\)
contact interaction is written as
\begin{equation}
V^{pp}_{\rm CT,full}
=
V^{I=1}_{\rm CT,sym.}
+
\Delta V^{pp}_{\rm CT},
\end{equation}
where \(V^{I=1}_{\rm CT,sym.}\) denotes the isospin-symmetric
contact interaction projected onto the \(I=1\) channel, and
\(\Delta V^{pp}_{\rm CT}\) represents the additional CSB
correction in the \(pp\) channel. The latter reads
\begin{widetext}
\begin{align}
\label{eq:LO_CT}
&\nonumber \Delta V^{pp}_{\rm CT}\l( \bm{p},\bm{p}'\r)=  C_S^{pp}\l[ \bar{u}\l( \bm{p},s_1\r) u\l( \bm{p}' ,s_1' \r) \r] \l[ \bar{u}\l( -\bm{p}' ,s_2' \r)  u \l( -\bm{p}, s_2 \r)\r]
+  C_V^{pp}\l[ \bar{u}\l( \bm{p},s_1\r)\gamma_\mu u\l( \bm{p}' ,s_1' \r) \r] \l[ \bar{u}\l( -\bm{p}' ,s_2' \r) \gamma^\mu u \l( -\bm{p}, s_2 \r)\r] \\
&+  C_{AV}^{pp}\l[ \bar{u}\l( \bm{p},s_1\r) \gamma_\mu\gamma_5 u\l( \bm{p}' ,s_1' \r) \r] \l[ \bar{u}\l( -\bm{p}' ,s_2' \r) \gamma^5\gamma_5 u \l( -\bm{p}, s_2 \r)\r] 
+  C_T^{pp}\l[ \bar{u}\l( \bm{p},s_1\r) \sigma_{\mu\nu} u\l( \bm{p}' ,s_1' \r) \r] \l[ \bar{u}\l( -\bm{p}' ,s_2' \r) \sigma^{\mu\nu} u \l( -\bm{p}, s_2 \r)\r],
\end{align}
\end{widetext}
where $C^{pp}_{S,V,AV,T}$ is the new low-energy constants related to the $pp$ interaction. With this convention, the exact isospin-symmetric limit is
obtained by switching off the CSB contact terms and the
electromagnetic interaction, namely
\begin{equation}
C^{pp}_{S}
=
C^{pp}_{V}
=
C^{pp}_{AV}
=
C^{pp}_{T}
=
0 .
\end{equation}
Therefore, in this limit,
\begin{equation}
V^{pp}_{\rm CT,full}
\rightarrow
V^{I=1}_{\rm CT,sym}.
\end{equation}
The same \(I=1\) isospin-symmetric contact interaction also
applies to the \(nn\) and isovector \(np\) channels. Hence,
in the exact isospin-symmetric limit and in the absence of
electromagnetic interactions, one has
\begin{equation}
V^{pp}_{\rm CT,full}
=
V^{nn}_{\rm CT,full}
=
V^{np,I=1}_{\rm CT,full}
=
V^{I=1}_{\rm CT,sym}.
\end{equation}
Here \(V^{np,I=1}_{\rm CT,full}\) denotes the isovector
component of the \(np\) interaction. The \(np\) system also
contains an \(I=0\) component, which is not related to the
\(pp\) or \(nn\) channels by isospin symmetry.}

Notice that, in principle, the charge-dependent two-nucleon 
potentials contain not only the \(np\) and \(pp\) interactions, 
but also the \(nn\) interaction. {  In the exact isospin-symmetric 
limit and in the absence of electromagnetic interactions, the 
\(nn\), \(pp\), and isovector \(np\) contact interactions reduce 
to the same \(I=1\) isospin-symmetric contact interaction. 
Possible charge-symmetry-breaking corrections in the \(nn\) 
channel could be introduced in analogy to \(\Delta V^{pp}_{\rm CT}\).} 
However, we do not construct the \(nn\) interaction in the present 
work because of the lack of experimental information on the 
\(nn\) scattering process, especially the scattering phase shifts.


\subsubsection{Isospin-breaking from the electromagnetic interaction}
The electromagnetic interaction also causes isospin-breaking effects due to the different charges of the up and down quarks.  In the chiral effective field theory, the most significant contribution from the electromagnetic interaction is the Coulomb force~\footnote{In practical calculation, the Coulomb force is obtained by a Fourier transformation of the Coulomb potential in the coordinate space integrated to radius $R$, whose explicit expression is shown in Eq.~(C.5) in Ref.~\cite{Epelbaum:2004fk}, for the sake of formulating $S$-matrix in terms of asymptotic Coulomb states. Here, we just want to show the explicit expression for the covariant Coulomb potential. } ( static $1\gamma$-exchange), 
\begin{widetext}
\begin{align}
V_{\text{Coulomb}} (\bm{p},\bm{p'})  = -e^2  \frac{\l[ \bar{u}\l( \bm{p},s_1 \r) \gamma_\mu  u\l( \bm{p}',s_1' \r) \r] \l[\bar{u}\l( -\bm{p}',s_2' \r)  \gamma_\mu u\l( -\bm{p},s_2 \r)\r] }{(E_p'-E_p)^2-(\bm{p'}-\bm{p})^2}.
\end{align}
\end{widetext}
In addition to the Coulomb force, the electromagnetic interaction also contains other corrections such as the ``improved Coulomb force" ($V_{\text{C2}}$), the magnetic moment interaction ($V_{\text{MM}}$), and the vacuum polarization potential ($V_{\text{VP}}$). However, these corrections are generally of higher order in the chiral expansion, and their contributions to the phase shifts are usually insignificant. So in the present calculation, we do not include them explicitly. We remind the reader that contributions from the aforementioned corrections may be indispensable for specific observables at certain momenta~\cite {Epelbaum:2004fk}. 

\section{Scattering equation and phase shifts}
\label{sec:Scat}
The partial wave projected scattering $T$-matrix is obtained by solving the Thompson equation in the $LSJ$ basis,
\begin{align}
\nonumber
    T^{SJ}_{L',L}\l(p',p \r)&=V^{SJ}_{L',L}\l(p',p \r) +\sum_{L''}\int_0^{+\infty} \frac{k^2 \text{d}k}{\l(2\pi\r)^3}V^{SJ}_{L',L}\l(p',k \r) \\\nonumber 
    & \times \frac{m_N^2}{2\l( k^2+m_N^2 \r)} \frac{1}{\sqrt{p^2+m_N^2}-\sqrt{k^2+m_N^2}+i\epsilon}\\
    & \times T^{SJ}_{L'',L} \l( k,p \r).
\end{align}

We separately solve the Thompson equation in the isospin basis for $I=0$ and $I=1$ ( only $I=1$ for the $pp$ process) and fit the resulting phase shifts to those in Ref.~\cite{Stoks:1993tb} to determine the corresponding LECs. To remove the ultraviolet divergences, the potential is regularized with a non-local Gaussian-type cut-off function,
\begin{align}\label{eq:cutoff}
    f^\Lambda \l( p,p' \r) = \exp\l[-\l( p^6+p'^6 \r)/\Lambda^6\r],
\end{align}
with the cut-off value  { $\Lambda = 700$ MeV.}
As already argued in Ref.~\cite{Xiao:2024jmu}, the employed regulator here is sort of ``old-fashioned" in the light of the studies on the $NN$ interactions in the conventional ChEFT because it distorts the long-range part of the potential, thus slowing down the convergence~\cite{Reinert:2017usi}. However, implementing a relatively modern regulator of the semi-local ~\cite{Reinert:2017usi} type in the present framework is not straightforward because it involves recalculating the covariant pion-exchange potential in the spectral functional renormalization (SFR) method and projecting it into the $LSJ$ basis. Thus, we leave it as future work. 

{  To consistently incorporate the Coulomb interaction within the Thompson equation, the Coulomb potential has to be modified with relativistic kinematic factors,
\begin{align}
    V^{\text{Thom.}}_{\text{Coulomb}} ({p},{p}';s)= \sqrt{\xi({p};s)} V_{\text{Coulomb}}({p},p') \sqrt{\xi({p}';s)},
\end{align}
with
\begin{align}
    \xi({k};s) = \frac{4 E_k^2}{m_N(\sqrt{s}+2E_k)},  ~~~~~~  E_k=\sqrt{k^2+m_N^2}.
\end{align}

This form follows from rewriting the Thompson equation as a Lippmann–Schwinger–type equation with an effective potential; see, e.g., Ref.~\cite{Holzenkamp:1989tq}.
}

The partial wave $S$ matrix is related to the on-shell 
$T$ matrix by,
\begin{align}
    S^{SJ}_{L',L}\l( p \r) = \delta_{L',L} - i \frac{ p~ M_N^2 }{8\pi^2 E_p} T^{SJ}_{L',L}\l(p \r).
\end{align}

Phase shifts and mixing angles can be obtained from the matrix $S$ using the idea of ``Stapp”~\cite{Stapp:1956mz}. For the single channels, 
\begin{align}
    \delta_L = \frac{1}{2} \arctan \frac{\text{Im}~S_L}{\text{Re}~S_L}.
\end{align}
For coupled channels, 
\begin{align}
   \delta_{J \pm 1} &= \frac{1}{2} \arctan \frac{\text{Im} ~\eta_{J \pm1}  }{\text{Re} ~\eta_{J \pm1}  } ,\\  \nonumber
  \epsilon_J &=   \frac{1}{2} \arctan  \frac{i \l( S_{J-1, J+1}+S_{J+1, J-1}\r)}{2\sqrt{S_{J-1, J-1} S_{J+1, J+1}}},
\end{align}
where $\eta_J =  \frac{S_{J,J}}{\cos 2 \epsilon_J}  $.

For the $pp$ process, the $S$-matrix must be formulated in terms of asymptotic Coulomb states due to the Coulomb interaction. This procedure is standard and has been discussed thoroughly; see Ref.~\cite{Vincent:1974zz} for more details.
For the uncoupled channel,
\begin{align}
    \tan \delta_L^l= \frac{\tan \delta_L^s[F, G_0]+ [F,F_0]}{[F_0,G]+\tan \delta_L^s [G_0,G]},
\end{align}
with $\delta^l_L/\delta_L^s$ refers to the phase shifts with/without the Coulomb interaction, $F(G)$ is the (ir)regular Coulomb function, $F_0(G_0)$ is the Bessel (Neumann) function, and 
\begin{align}
    [F,G]=\l( G \frac{d F}{d r}-F\frac{dG}{dr} \r)_{r=R},
\end{align}
where $R$ is the matching radius and $R=12$ fm in this work as suggested in Ref.~\cite{Epelbaum:2004fk}.

For the coupled case, $F(r)$ and $G(r)$ are $2 \times 2$ matrix of the following form,
\begin{align}
    F(r)= \l(
    \begin{matrix}
        F_{J-1}(r)&0\\
        0& F_{J+1} (r)
    \end{matrix}
    \r),~
    G(r)= \l(
    \begin{matrix}
        G_{J-1}(r)&0\\
        0& G_{J+1} (r)
    \end{matrix}
    \r).
\end{align}
The $S$-matrix with the Coulomb interaction $S^l$ is related with that without the Coulomb interaction $S^s$ as 
\begin{align}
    S^l=\frac{U^{-} M_1^{-1} M_2- u^-}{U^+-u^+ M_1^{-1} M_2},
\end{align}
where 
\begin{align}
    M_1&=U_0^- + S^s U_0^+ \\\nonumber
    M_2& = u_0^- + S^s u_0^+ ,
\end{align}
and 
\begin{align}
    U^{\pm}_{(0)}= \l[ F_{(0)}(r)\mp i G_{(0)}(r) \r]_{r=R},\quad u^{\pm}_{(0)} = \frac{d  U^\pm_{(0)}}{dr}.
\end{align}
Since the above matching procedure only involves the $S$-matrix and Coulomb functions, it is scattering equation-independent as long as one uses the ``Stapp" parameterization to calculate the phase shifts.

\section{Results and Discussions}
\label{sec:Results}
\subsection{Numerical details}
In this section, we provide a detailed introduction to the parameters and fitting strategies employed in this work. 

The nucleon mass $M_N = 939$ MeV, the pion masses $M_{\pi^0} = 135$ MeV and $M_{\pi^\pm} = 139$ MeV~\cite{ParticleDataGroup:2024cfk}. The pion decay constant $f_\pi = 92.4 $ MeV,  the axial coupling constant $g_A = 1.29$~\cite{Machleidt:2011zz}.
{ The pion-nucleon low-energy constants $c_{1,2,3,4}$ are essential inputs for the two-pion exchange (TPE) contributions. It is widely recognized that these LECs can be obtained using the Roy-Steiner equation~\cite {Hoferichter:2015tha}. However, the values of these LECs can be scheme-dependent. To maintain theoretical consistency within our relativistic framework, we use $ c_1 =-1.39$, $c_2=4.01$, $c_3=-6.61$, $c_4=3.92$, all in units of GeV$^{-1}$, determined from a fit to $\pi N$ scattering data within covariant chiral EFT~\cite{Chen:2012nx}, as these have been successfully applied to describe the isospin-symmetric nuclear force. This choice ensures that the short-range dynamics and the relativistic loop contributions are treated under a unified power-counting and regularization scheme.}

There are in total 19 LECs for the isospin-conserving interaction and $4$ additional LECs for the isospin-breaking interaction up to NNLO which are all determined by a simultaneous fit to the $J\leq2$ $np$ and $pp$ scattering phase shifts~\cite{Stoks:1993tb} for the laboratory energy at $T_{\text{lab}} = 1, 5, 10, 25, 50, 100, { {150,}} 200$ MeV.  The operator structures of the isospin-conserving contact interactions up to NNLO are the same as those given in our previous work~\cite{Lu:2021gsb}, and are therefore not repeated here. The LECs for the isospin-conserving operators $O_{1 - 17}$ and $D_{1-2}$, and isospin-breaking operators $C_{S,V,AV,T}^{pp}$,   and { $\chi^2_{\text{eff}}$/datum with $\chi^2_\text{eff}=\sum_i \left(\frac{\delta^i - \delta^i_{\text{PWA93}}}{ \Delta^i}\right)^2$} are shown in Table~\ref{tb:LECs} and Table~\ref{tb:chi2} respectively. The {full $pp$-channel effective tensor coefficient
\(C_{T,\mathrm{full}}^{pp}=O_4+C_T^{pp}\)} is fixed at zero because the four additional $pp$ LECs contribute only to the $J\leq1$ partial waves, while only 3 $pp$ partial waves exist for $J\leq1$. The calculated $\chi^2_{\text{eff}}/\text{datum}$ for the LO results is not shown because isospin-breaking effects start at NLO, and the fitting strategy used at LO differs from those used at NLO and NNLO, as explained. 
{ To ensure a meaningful definition of $\chi^2_\text{eff}$, we adopt the methodology proposed in Ref.~\cite{Epelbaum:2014efa}, the uncertainty for a given phase shift (or mixing angle) $\delta$ in the channel $i$ at a given energy is defined as,
\begin{align}
    \Delta^i = \text{max}  \left(~ \Delta^i_{\text{PWA93}},~ |\delta^i_{\text{models}}-\delta^i_{\text{PWA93}}|  ~\right),
\end{align}
where $\Delta^i_{\text{PWA93}}$ refers to the statistical error of partial wave analysis in the channel $i$~\cite{Stoks:1993tb}, and $\text{models} \in \left\{  \text{NijmI, NijmII, Reid93}  \right\}$, the phase shifts for all potential models are sourced from NN-Online.~\cite{NNOnline}. 

In practice, the uncertainties obtained in this way can become unrealistically small for certain data points, leading to an overconstrained fit. To avoid an excessive weighting of such points in the $\chi^2_{\text{eff}}$ function, we impose a lower bound on the uncertainties according to

\begin{align}
    \Delta^i \;\to\; \max \left( \Delta^i,~ \Delta_{\text{min}}^i \right), ~ 
    \Delta_{\text{min}}^i = \min \left( 0.1^\circ,~ 0.1\,|\delta^i_{\text{PWA93}}| \right).
\end{align}

We emphasize that the $\chi^2_{\text{eff}}$ defined in the present work measures the deviation of the calculated phase shifts from those of the PWA93 analysis rather than the deviation from the experimental $NN$ scattering data directly. Consequently, the resulting $\chi^2_{\text{eff}}$/datum should not be compared directly with the conventional benchmark values quoted for high-precision $NN$ potentials, but should instead be interpreted as a measure of the level of agreement with the partial-wave analysis.}

\begin{table}[htb]
\centering
\caption{LECs in units of $10^4~{\rm GeV}^{-2}$ for the covariant LO, NLO, and NNLO results.
The covariant contact operator structures corresponding to $O_{1-17}$ are denoted as $\tilde{O}_{1-17}$ and listed in Table~III of Ref.~\cite{Xiao:2018jot}.
The two additional operators associated with $D_1$ and $D_2$ are defined in Eq.~(2) of the Supplemental Material of Ref.~\cite{Lu:2021gsb}.}
\label{tb:LECs}
\begin{tabular}{lrrr}
\hline
\hline
Operator & LO & NLO & NNLO \\
\hline
$O_1$         & $-1.23$ & $-13.51$ & $7.77$ \\
$O_2$         & $-0.22$ & $-3.08$  & $0.52$ \\
$O_3$         & $-0.88$ & $3.65$   & $-4.98$ \\
$O_4$         & $0.29$  & $-0.70$  & $0.43$ \\
$O_5$         & --      & $10.17$  & $3.38$ \\
$O_6$         & --      & $-3.96$  & $-0.01$ \\
$O_7$         & --      & $-0.42$  & $-4.05$ \\
$O_8$         & --      & $0.49$   & $2.55$ \\
$O_9$         & --      & $-1.61$  & $3.69$ \\
$O_{10}$      & --      & $1.55$   & $-2.00$ \\
$O_{11}$      & --      & $3.02$   & $1.28$ \\
$O_{12}$      & --      & $1.98$   & $3.23$ \\
$O_{13}$      & --      & $0.55$   & $-3.04$ \\
$O_{14}$      & --      & $-4.43$  & $-2.82$ \\
$O_{15}$      & --      & $3.84$   & $2.06$ \\
$O_{16}$      & --      & $4.42$   & $-1.24$ \\
$O_{17}$      & --      & $-0.27$  & $2.06$ \\
$D_1$         & --      & $0.46$   & $-1.99$ \\
$D_2$         & --      & $0.28$   & $-1.17$ \\
$C_S^{pp}$    & --      & --        & $-10.82$ \\
$C_V^{pp}$    & --      & --        & $3.92$ \\
$C_{AV}^{pp}$ & --      & --        & $10.23$ \\
$C_T^{pp}$    & --      & --        & $-0.43$ \\
\hline
\hline
\end{tabular}
\end{table}

\begin{table}[ht]
\centering
\caption{$\chi^2_{\rm eff}/{\rm datum}$ at different orders for the $J\leq 2$ partial waves.
``$np$'' and ``$pp$'' refer to the total $\chi^2_{\rm eff}/{\rm datum}$ for the $np$ and $pp$ processes, respectively.
The partial waves listed below the $np$ and $pp$ entries correspond to the $np$ and $pp$ scattering channels.}
\label{tb:chi2}
\begin{tabular}{lrr}
\hline
\hline
Quantity/channel & NLO & NNLO \\
\hline
Total & $31.50$ & $7.51$ \\
\hline
$np$ total & $30.36$ & $4.53$ \\
$^{1}S_0$  & $14.83$ & $1.33$ \\
$^{3}P_0$  & $3.24$  & $16.74$ \\
$^{1}P_1$  & $0.80$  & $1.30$ \\
$^{3}P_1$  & $1.84$  & $26.63$ \\
$^{1}D_2$  & $22.79$ & $0.83$ \\
$^{3}D_2$  & $4.73$  & $1.43$ \\
$^{3}S_1$  & $63.37$ & $0.78$ \\
$^{3}D_1$  & $239.74$ & $0.36$ \\
$\epsilon_1$ & $2.00$ & $1.53$ \\
$^{3}P_2$  & $10.10$ & $2.63$ \\
$^{3}F_2$  & $0.69$  & $0.69$ \\
$\epsilon_2$ & $0.16$ & $0.08$ \\
\hline
$pp$ total & $33.46$ & $12.62$ \\
$^{1}S_0$  & $181.83$ & $11.41$ \\
$^{3}P_0$  & $11.58$ & $21.07$ \\
$^{3}P_1$  & $2.33$  & $48.81$ \\
$^{1}D_2$  & $22.15$ & $1.74$ \\
$^{3}P_2$  & $15.54$ & $4.44$ \\
$^{3}F_2$  & $0.20$  & $0.09$ \\
$\epsilon_2$ & $0.57$ & $0.76$ \\
\hline
\hline
\end{tabular}
\end{table}

Notice that isospin-breaking effects start at NLO, so the LO LECs are only fit to the $np$ phase shifts at laboratory energies below 100 MeV. Their values differ from those reported in Ref.~\cite{Lu:2021gsb} due to a different choice of regulators.

{It is worth noting that the interpretation of the natural size of fitted LECs is highly dependent on the specific operator definitions, normalization conventions, and the regularization scheme employed. In
the covariant chiral EFT framework, the power counting and the relativistic structure of the contact operators differ from their non-relativistic counterparts. Consequently, the benchmark scales from naturalness established in non-relativistic formulations may not
directly apply here.

In particular, the values of \(C^{pp}_{S,V,AV,T}\), which denote the coefficients of the additional $pp$-channel correction \(\Delta V_{CT}^{pp}\), do not show a simple numerical suppression.
However, individual fitted LECs and the potential kernel are representation- and regulator-dependent quantities and are not observables, and the contact interactions are iterated nonperturbatively in the Thompson equation.

To further illustrate the impact of the $pp$-channel short-range correction on phase shifts, we performed an additional diagnostic calculation. We switched off the additional $pp$-channel correction by setting \(C^{pp}_{S,V,AV,T}=0\), while keeping all other $pp$ ingredients unchanged, including the Coulomb interaction, the pion-mass splitting in OPE, and the Coulomb matching procedure.  We define
\begin{equation}
\Delta\delta_{CT}^{pp}=\delta^{pp}-\delta_{C_{S,V,AV,T}^{pp}=0}^{pp}.
\end{equation}
The resulting phase-shift changes for the most affected channels are listed in Table~\ref{tab:delta_ct_pp}. The changes in the \(^{3}P_1\) phase shifts are smaller than \(0.01^\circ\) at all fitted energies and are therefore not shown.

\begin{table}[t]
\caption{ Phase shifts changes (in units of degrees) in the $pp$ phase shifts induced by switching off the additional $pp$-channel correction \(C^{pp}_{S,V,AV,T}\), while keeping all other $pp$
ingredients unchanged.}
\label{tab:delta_ct_pp}
\resizebox{\columnwidth}{!}{
\begin{tabular}{c|cccccccc}
\hline
\hline
\(T_{\rm lab}\) (MeV) & 1 & 5 & 10 & 25 & 50 & 100 & 150 & 200 \\
\hline
$\Delta\delta_{CT}^{pp} (^{1}S_0)$ & $-5.885$ & $-4.881$ & $-3.887$ & $-3.125$ & $-2.955$ & $-3.110$ & $-3.299$ & $-3.414$ \\
$\Delta\delta_{CT}^{pp} (^{3}P_0)$ & $-0.001$ & $-0.014$ & $-0.041$ & $-0.148$ & $-0.298$ & $-0.465$ & $-0.658$ & 2.775 \\
\hline
\hline
\end{tabular}
}
\end{table}

Table~\ref{tab:delta_ct_pp} shows that the effect of the $pp$-channel short-range correction is mainly in the \(^{1}S_0\) channel. The changes in the \(^{3}P_0\) phase shift are moderate over
most of the fitted region, while the \(^{3}P_1\) phase shift is practically unaffected. The somewhat larger value in the \(^{3}P_0\) channel at \(200~\mathrm{MeV}\) occurs only at the upper end of the fitted energy range and should not be interpreted as a uniform enhancement of the
$pp$-channel short-range correction in this partial wave. This diagnostic demonstrates that the numerical size of the fitted contact correction should not be interpreted as a direct measure of its impact in each physical partial wave.}

To estimate the theoretical uncertainty, we closely follow the Bayesian method proposed in Refs.~\cite{Furnstahl:2015rha,Melendez:2017phj,Melendez:2019izc}. { The specific implementation follows that of our previous work~\cite{Lu:2021gsb}.} Since the isospin-breaking effects start from NLO, we use  LO $np$ phase shifts as input to calculate the NLO $pp$ phase shifts uncertainty.

\subsection{Phase shifts}

\begin{figure*}[htb]
\centering
\includegraphics[width=1.0\textwidth]{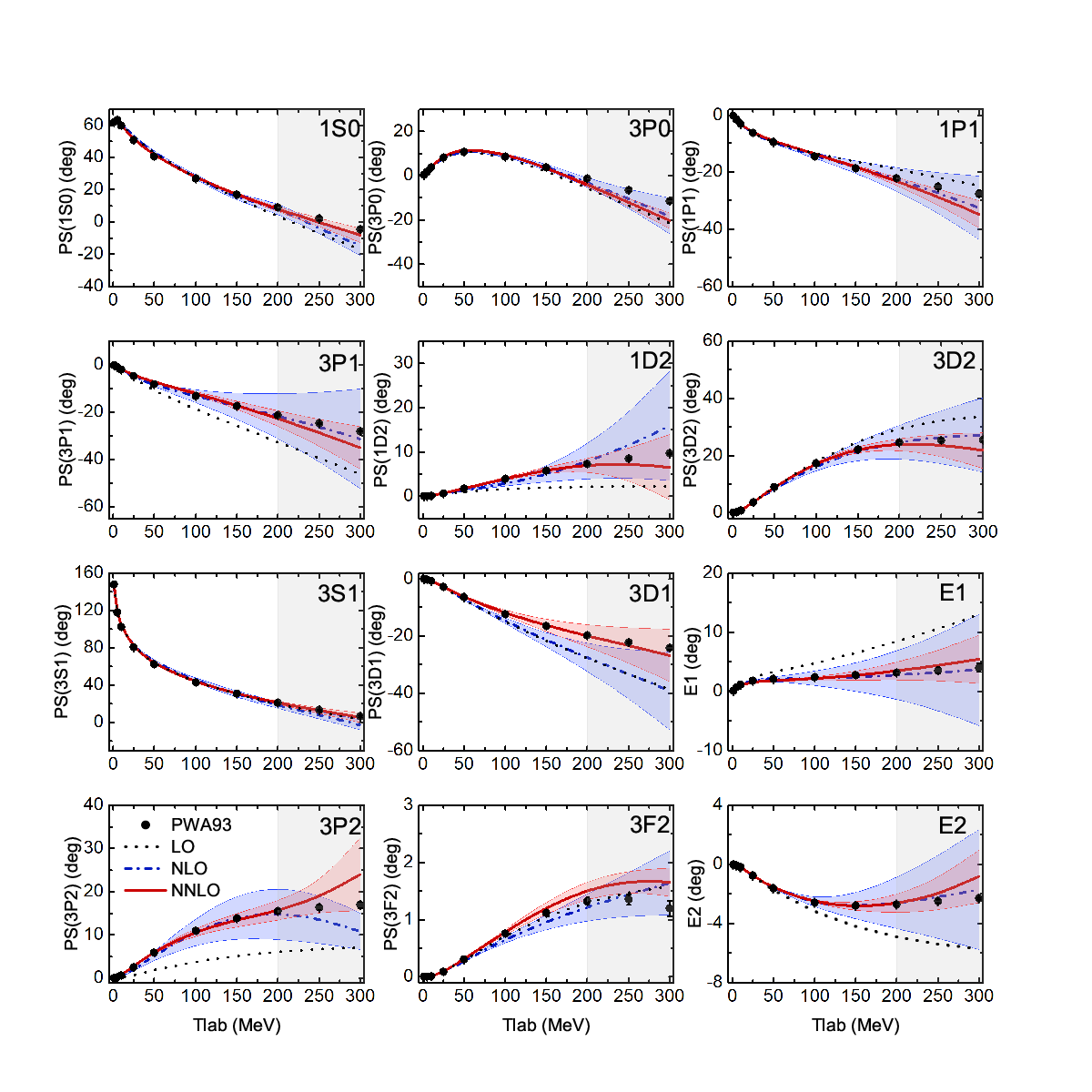}
\caption{ {  Phase shifts (PS) as a function of laboratory energy (Tlab) for $np$ scattering for the $J\leq2$ partial waves.
The black dot, blue dot-dashed, and red solid lines correspond to LO, NLO, and NNLO results, respectively.The heavy dots represent the PWA93 phase shifts~\cite{Stoks:1993tb}, 
with the associated uncertainties indicated by the error bars as discussed in the text.
The blue and red shaded bands indicate the estimated theoretical uncertainties using the Bayesian method~\cite{Furnstahl:2015rha,Melendez:2017phj,Melendez:2019izc}. The gray band shows the PS in the energy region not included in the fit, where the EFT results are predictions.} }
\label{fig:np}
\end{figure*}

\begin{figure*}[htbp]
\centering
\includegraphics[width=1.0\textwidth]{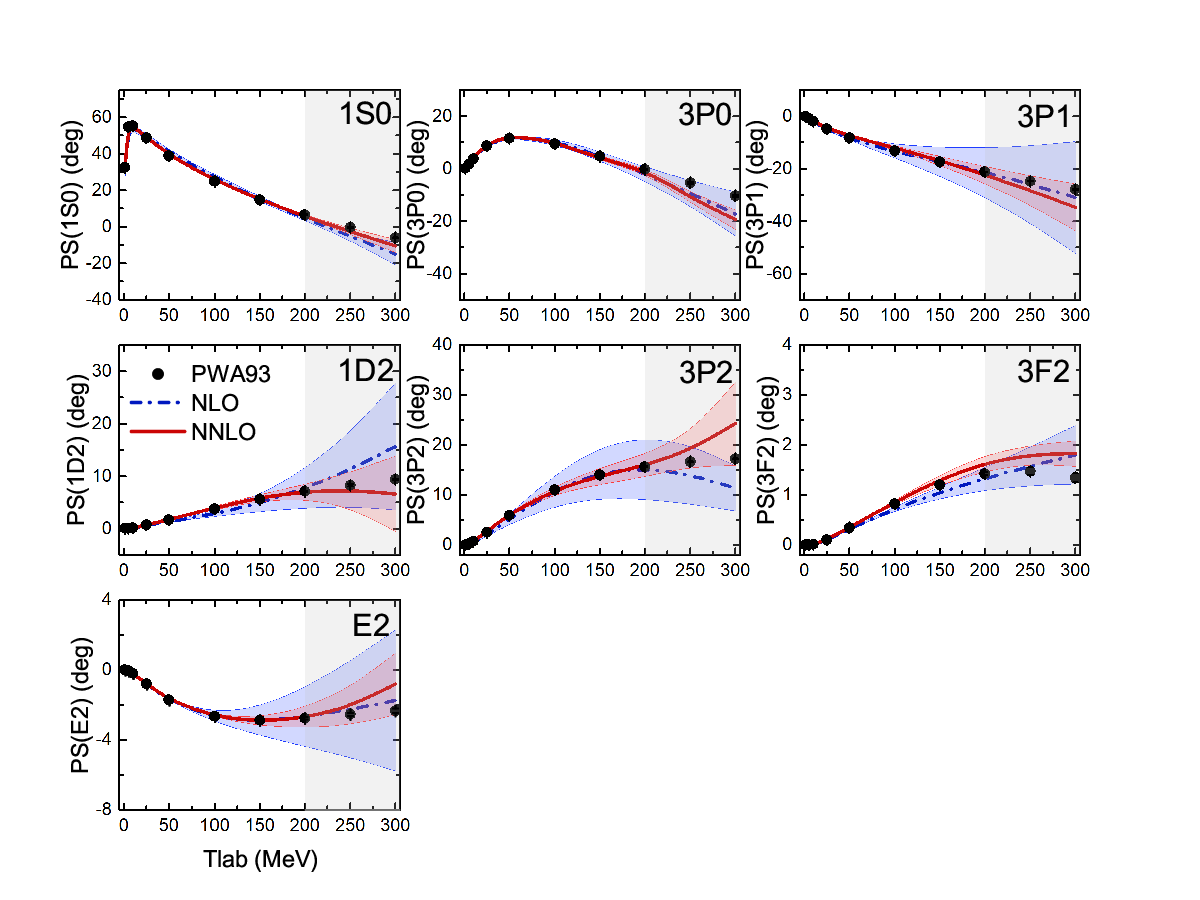}
\caption{ The $pp$ phase shift for the $J\leq2$ partial waves. { The notation is the same as in Fig.~\ref{fig:np}.} }
\label{fig:pp}
\end{figure*}

Fig.~\ref{fig:np} and Fig.~\ref{fig:pp} show the $np$ and $pp$ scattering phase shifts for $J\leq2$ partial waves as a function of laboratory energy, calculated in covariant chiral effective field theory. 

For the $np$ scattering, 
{ the calculated NNLO phase shifts agree well with the PWA93 results in the fitted energy region as in Ref.~\cite{Lu:2021gsb}. At NLO, the description of the $^1S_0$ phase shift becomes slightly worse, as no additional isospin-breaking LECs are introduced at this order. Compared to our previous study~\cite{Lu:2021gsb}, the $\chi^2_{\text{eff}}/\text{datum}$ at NLO in the coupled $^3S_1$–$^3D_1$ channel is somewhat larger. This can be attributed to this channel's greater sensitivity to cutoff artifacts. In Ref.~\cite{Lu:2021gsb}, a sharp cutoff with $\Lambda=900$ MeV was employed, while in the present work we adopt a Gaussian regulator with a lower cutoff scale $\Lambda=700$ MeV. As a result, high-momentum components are suppressed more strongly, which leads to visible differences in the phase shifts, especially at intermediate and higher energies, and consequently to a somewhat larger $\chi^2_{\text{eff}}/\text{datum}$ at NLO.
At NNLO, the description improves significantly even though no additional low-energy constants are introduced. This improvement mainly originates from the inclusion of subleading two-pion-exchange contributions, which provide a more accurate description of the intermediate-range interaction. These higher-order contributions partially compensate for the regulator dependence present at NLO and lead to a better reproduction of the phase shifts in the $^3S_1$–$^3D_1$ channel, thereby reducing the $\chi^2_{\text{eff}}/\text{datum}$. This behavior is consistent with the expected order-by-order improvement pattern of chiral effective field theory.} In the higher energy region, although the description of the PWA93 phase shifts is less accurate than in the lower energy region, they still fall within the uncertainties of the NNLO results. This is to be expected, since EFT is known to work more effectively at low energies. Compared with the results shown in Ref.~\cite{Lu:2021gsb}, the LO results appear to exhibit different behavior, which is related to the new regulator adopted in this work.

For the $pp$ scattering, { {overall, the calculated $pp$ phase shifts at NLO and NNLO show a behavior similar to that observed in the $np$ case: the NLO and NNLO results are already close to the PWA93 phase shifts at low energies, while at higher energies most of the PWA93 phase shifts remain within the estimated theoretical uncertainties of the EFT framework. The majority of the $\chi^2_{\text{eff}}/\text{datum}$ }} for the NLO results comes from the $^1S_0$ partial wave because the $pp$ and $np$ scattering phase shifts show relatively large discrepancies at low energy, and no additional $pp$ contact terms at this order. This situation improves at NNLO due to the presence of four $pp$ contact terms. For the higher energy regime, in general, the NLO and NNLO results agree with the PWA93 phase shifts, { while some deviations are observed in the $J \leq 2$ coupled channels, which can be attributed to the relatively large contributions from the NNLO TPE, as also discussed in Ref.~\cite{Lu:2021gsb}. }

{ It is important to note that while the numerical impact of relativistic corrections on low-energy isospin-breaking observables may appear perturbative, the primary value of the present formulation lies in its structural integrity. By treating the $pp$  and $np$ systems on an equal footing within a Lorentz-invariant framework, we provide a robust foundation for extending chiral nuclear forces to higher energies and denser systems, such as the study of the symmetry energy in isospin-asymmetric nuclear matter. The satisfactory agreement with the PWA93 data (as shown in Figs.~\ref{fig:np} and ~\ref{fig:pp}) confirms that our systematic inclusion of isospin breaking contact terms at NNLO successfully compensates for the inherent differences in the $pp$ and $np$ sectors within this covariant scheme.}

\section{ Summary and Outlook}
\label{sec:summary}
Based on covariant chiral effective field theory, we extend the 
relativistic chiral $NN$ interaction to the proton-proton system, and construct the charge-dependent covariant chiral $NN$ interactions up to NNLO. The isospin-breaking effects, including pion mass splitting, Coulomb interactions, and charge-dependent contact terms, are considered systematically. All low-energy constants are determined by fitting to the $np$ and $pp$ scattering phase shifts simultaneously. The calculated phase shifts show a satisfactory agreement with the PWA93 phase shifts. 

The charge-dependent $NN$ interactions obtained in this work may serve as a useful input for future studies of isospin-breaking phenomena in nuclear systems within the covariant chiral EFT framework, such as disentangling charge-symmetry and charge-independence breaking effects in $NN$ scattering, resolving mirror energy differences,  deepening the understanding of charge-exchange processes and electroweak transitions, and determining the symmetry energy and the equation of state of isospin-asymmetric matter.

\section*{Data Availability}
The numerical data underlying Figs.~\ref{fig:np} and \ref{fig:pp} are provided as separate data files with this article.

\section{Acknowledgments}
Yang Xiao thanks Prof. Hao-Zhao Liang for useful discussions. This work was supported in part by the National Natural Science Foundation of China under Grants No.12347113,  No.12505096, No.1243007, the Chinese Postdoctoral Science Foundation under Grants No.2022M720360, and the Institute for Basic Science (IBS-R031-D1). Yang Xiao and Jun-Xu Lu thank the Fundamental Research Funds for Central Universities for the support. Yang Xiao thanks the support from the China Scholarship Council. 
\bibliography{Refs}

\end{document}